\documentclass[aps,prc,twocolumn,unsortedaddress,superscriptaddress,nobibnotes]{revtex4}
\usepackage{color,graphicx}
\usepackage{array,amsmath,amssymb,pifont,mathrsfs,latexsym}

{\begin{list}{}{%
      \settowidth{\labelwidth}{\textbf{#1}}%
      \setlength{\leftmargin}{\labelwidth}
      \addtolength{\leftmargin}{0.5em}}}
{\end{list}}
\newcommand{\nuc}[2]{\ensuremath{^{\text{#1}}\text{#2}}}
\newcommand{\mevc}{\mbox{MeV\hspace{-0.111em}/$c$}}

\newcommand{\meva}{\mbox{MeV\hspace{-0.111em}/\textit{A}}}

\newcommand{\CaFCaF}{\nuc{40}{Ca}+\nuc{40}{Ca}}
\newcommand{\CaFECaFE}{\nuc{48}{Ca}+\nuc{48}{Ca}}
\newcommand{\pp}{$p$-$p$ }
\newcommand{\np}{$n$-$p$ }
\newcommand{\nn}{$n$-$n$ }

\begin{document}

\title{Angular Dependence in Proton-Proton Correlation Functions in Central \nuc{40}{Ca}+\nuc{40}{Ca} and \nuc{48}{Ca}+\nuc{48}{Ca} Reactions}
\author{V. Henzl}
\altaffiliation{Current address: Los Alamos National Laboratory, Los Alamos, NM 87545, USA}
\affiliation{National Superconducting Cyclotron Laboratory, Michigan State University, East Lansing, MI 48864, USA}
\author{M. A. Kilburn}
\affiliation{National Superconducting Cyclotron Laboratory, Michigan State University, East Lansing, MI 48864, USA}
\affiliation{Department of Physics and Astronomy, Michigan State University, East Lansing, MI 48864, USA}
\author{Z. Chaj\c{e}cki}
\affiliation{National Superconducting Cyclotron Laboratory, Michigan State University, East Lansing, MI 48864, USA}
\author{D. Henzlova}
\altaffiliation{Current address: Los Alamos National Laboratory, Los Alamos, NM 87545, USA}
\affiliation{National Superconducting Cyclotron Laboratory, Michigan State University, East Lansing, MI 48864, USA}
\author{W. G. Lynch}
\email[Email comments to: ]{lynch@nscl.msu.edu}
\affiliation{National Superconducting Cyclotron Laboratory, Michigan State University, East Lansing, MI 48864, USA}
\affiliation{Department of Physics and Astronomy, Michigan State University, East Lansing, MI 48864, USA}
\author{D. Brown}
\affiliation{National Superconducting Cyclotron Laboratory, Michigan State University, East Lansing, MI 48864, USA}
\affiliation{Department of Physics and Astronomy, Michigan State University, East Lansing, MI 48864, USA}
\author{A. Chbihi}
\affiliation{GANIL, CEA et IN2P3/CNRS, F-14076 Caen, France}
\author{D. Coupland}
\affiliation{National Superconducting Cyclotron Laboratory, Michigan State University, East Lansing, MI 48864, USA}
\affiliation{Department of Physics and Astronomy, Michigan State University, East Lansing, MI 48864, USA}
\author{P. Danielewicz}
\affiliation{National Superconducting Cyclotron Laboratory, Michigan State University, East Lansing, MI 48864, USA}
\affiliation{Department of Physics and Astronomy, Michigan State University, East Lansing, MI 48864, USA}
\author{R. deSouza}
\affiliation{Department of Chemistry, Indiana University, Bloomington, IN 47405, USA}
\author{M. Famiano}
\affiliation{Department of Physics, Western Michigan University, Kalamazoo, MI 49008, USA}
\author{C. Herlitzius}
\affiliation{National Superconducting Cyclotron Laboratory, Michigan State University, East Lansing, MI 48864, USA}
\affiliation{Joint Institute of Nuclear Astrophysics, Michigan State University, East Lansing, MI 48864, USA}
\author{S. Hudan}
\affiliation{Department of Chemistry, Indiana University, Bloomington, IN 47405, USA}
\author{Jenny Lee}
\affiliation{National Superconducting Cyclotron Laboratory, Michigan State University, East Lansing, MI 48864, USA}
\affiliation{Department of Physics and Astronomy, Michigan State University, East Lansing, MI 48864, USA}
\author{S. Lukyanov}
\affiliation{FLNR, JINR, 141980 Dubna, Moscow region, Russian Federation}
\author{A. M. Rogers}
\altaffiliation{Current address: Physics Division, Argonne National Laboratory, Argonne, Illinois, 60439 USA}
\affiliation{National Superconducting Cyclotron Laboratory, Michigan State University, East Lansing, MI 48864, USA}
\affiliation{Department of Physics and Astronomy, Michigan State University, East Lansing, MI 48864, USA}
\author{A. Sanetullaev}
\affiliation{National Superconducting Cyclotron Laboratory, Michigan State University, East Lansing, MI 48864, USA}
\affiliation{Department of Physics and Astronomy, Michigan State University, East Lansing, MI 48864, USA}
\author{L. Sobotka}
\affiliation{Department of Chemistry, Washington University, St. Louis, MO 63130, USA}
\author{Z. Y. Sun}
\affiliation{National Superconducting Cyclotron Laboratory, Michigan State University, East Lansing, MI 48864, USA}
\affiliation{Institute of Modern Physics, CAS, Lanzhou 730000, Peoples Republic of China}
\author{M. B. Tsang}
\affiliation{National Superconducting Cyclotron Laboratory, Michigan State University, East Lansing, MI 48864, USA}
\author{A. Vander Molen}
\affiliation{National Superconducting Cyclotron Laboratory, Michigan State University, East Lansing, MI 48864, USA}
\author{G. Verde}
\affiliation{INFN, Laboratori Nazionali del Sud, Catania, Italy}
\author{M. Wallace}
\altaffiliation{Current address: Los Alamos National Laboratory, Los Alamos, NM 87545, USA}
\affiliation{National Superconducting Cyclotron Laboratory, Michigan State University, East Lansing, MI 48864, USA}
\affiliation{Department of Physics and Astronomy, Michigan State University, East Lansing, MI 48864, USA}
\author{M. Youngs}
\affiliation{National Superconducting Cyclotron Laboratory, Michigan State University, East Lansing, MI 48864, USA}
\affiliation{Department of Physics and Astronomy, Michigan State University, East Lansing, MI 48864, USA}

\date{\today}

\begin{abstract}
  The angular dependence of proton-proton correlation functions is studied in central \nuc{40}{Ca}+\nuc{40}{Ca} and \nuc{48}{Ca}+\nuc{48}{Ca} nuclear reactions at E = 80 \meva. Measurements were performed with the HiRA detector complemented by the 4$\pi$ Array at NSCL.  A striking angular dependence in the laboratory frame is found within \pp correlation functions for both systems that greatly exceeds the measured and expected isospin dependent difference between the neutron-rich and neutron-deficient systems.
 Sources measured at backward angles reflect the participant zone of the reaction, while much larger sources observed at forward angles reflect the expanding, fragmenting and evaporating projectile remnants.  The decrease of the size of the source with increasing momentum is observed at backward angles while a weaker trend in the opposite direction is observed at forward angles.  The results are compared to the theoretical calculations using the BUU transport model.
\end{abstract}
\maketitle


\section{Introduction}

The spectra of particles emitted in nuclear reactions can include contributions from a variety of dynamical and statistical mechanisms characterized by vastly different timescales. Dynamical emission typically occurs over timescales as short as $10^{-22}$ seconds. Statistical emission can extend to much longer times. The descriptions of dynamical and statistical emission mechanisms require completely different theoretical formalisms. This complicates theoretical interpretations of measured spectra, as most experimental observables do not allow a model independent distinction between earlier dynamical and later statistical emission.

The correlation functions relevant to intensity interferometry investigation ~\cite{Hanbury:1954wr,Goldhaber:1960sf,boal90}, however, do not suffer this limitation. Due to their ability to  probe the space-time extent of the sources of emission, two particle correlation functions allow a distinction between early dynamical and later statistical emission. This has been used to probe the emission mechanisms of a variety of different particle types for a number of different reactions studied over a wide range of collision energies~\cite{zhu91,gaff95,handzy94,chen87}.

The sensitivity of the two proton correlation functions to the space-time extent of the source arises from the mutual nuclear (attractive) and Coulomb (repulsive) interactions between the two protons and from the antisymmetrized nature of their wave functions ~\cite{boal90}. Gates on the proton pair velocity provide information about the sources of these protons at different times during the reaction. Proton pairs with higher total momenta in the rest frame of the source preferentially reflect the space-time extent of that source at earlier emission times when the source is smaller. Smaller sources typically display larger and broader correlation functions~\cite{zhu91}. In contrast, proton pairs with lower total momenta tend to be emitted at later times after the source has expanded and cooled. Such sources typically display narrower, and weaker correlation functions.  Thus, correlation functions can track the time evolution of a cooling, expanding source.

Transport models~\cite{Chen:2003wp,Chen:2004kj} have revealed the existence of a sensitivity of two-nucleon (\pp, \np and \nn) correlation functions to the density dependence of the symmetry energy and some sensitivity to isospin in two-particle correlation functions have been observed~\cite{Ghetti:2003pv}. Physically, this sensitivity was shown to come from the effect of the symmetry energy on proton and neutron  potentials and their influence on the emission times of particles during the pre-equilibrium stages of the collision~\cite{Chen:2003wp}. This suggests that investigations of isospin effects on reaction dynamics and their links to the density dependence of the symmetry energy~\cite{Li:2008gp}, may profit from a more clear understanding of the time characteristics of different particle emission processes and by the capability of isolating emissions from the early pre-equilibrium stages of the reaction~\cite{Chen:2003wp,Chen:2004kj}.

Stimulated by these ideas, we have measured  \pp correlations over a wide angular and kinematic range with high statistics. In order to investigate the existence of isospin effects, we have compared results from reaction systems with different N/Z asymmetries, i.e. \nuc{40}{Ca}+\nuc{40}{Ca} (N/Z =1) and \nuc{48}{Ca}+\nuc{48}{Ca} (N/Z =1.4), at beam energies E/A=80 MeV.

Studies have been performed for the \nuc{36}{Ar}+\nuc{45}{Sc} reaction at E/A=80 MeV, providing some guidance for the dependence of the source on the momenta of the outgoing protons ~\cite{Lisa:1993xh,Lisa:1993zz}. The correlation functions measured in \nuc{36}{Ar}+\nuc{45}{Sc} reactions show a strong decrease in the source size with proton momentum, consistent with emission from an expanding and cooling participant source. BUU transport calculations generally reproduce these experimental trends.

 In our investigations of $\nuc{40}{Ca}+\nuc{40}{Ca}$ and $\nuc{48}{Ca}+\nuc{48}{Ca}$ collisions, we have measured the correlation functions over a broader range of angles than previous measurements, and have studied in detail the momentum dependent two-proton correlation functions at different angles. The obtained results show that the applied momentum gates have strikingly different effects on the size of sources corresponding to particles emitted at forward angles as compared to those detected at backward angles in the laboratory frame. The measurements show a strong influence of emission from the expanding, fragmenting and cooling spectator matter that was not evident in previous measurements. We also extract the fraction of protons emitted over short timescales during the collisions from the  height of the correlation function and the integral of the imaged source distribution. We see surprisingly little sensitivity of these fractions to the angle or momentum of the measured protons. In order to better distinguish dynamical from statistical emission mechanisms,  we also compared the extracted results to expectations of a BUU transport model~\cite{dan91}.

\section{Two Particle Correlations}

The two-proton correlation function probes the spacial and temporal information about the particle
emitting source because the magnitude of the final-state interactions and anti-symmetrization effect
depend on both the spacial separation and the relative momentum of particles~\cite{verde06}.

Theoretically, the correlation function is related to the space-time extent of the source by
the angle-averaged Koonin-Pratt equation~\cite{koon77, pratt90}
\begin{equation}
  \label{eq_kp}
  C(q)=1+R(q)=1+4\pi \int K(q,r)S(r)r^{2}dr,
\end{equation}
where
the two-particle source function, \textit{S(r)}, is the probability of emitting two protons with a spatial separation \textit{r}. In general protons are not emitted simultaneously. Then \textit{r} in Eq.~\ref{eq_kp} refers to the separation at the time the second proton is emitted. The source function satisfies the following normalization condition
\begin{equation}
\label{eq_source_integral}
4\pi \int S(r)r^2 dr = 1
\end{equation}
The angle-averaged kernel $K(q,r)$ is given by
\begin{equation}
  K(q,r)=|\phi_{q}(r)|^{2}-1.
  \label{eq_kernel}
\end{equation}
where $\phi_{q}(r)$ is the two proton
wave function measured at the separation distance between particles $r$ and at pair relative momentum $q$, defined in the center of mass of the pair by:
\begin{equation}
  q=|\vec{q}|=\frac{1}{2}|\vec{p_{1}}-\vec{p_{2}}|
  \label{eq_q}
\end{equation}

Within this approach, the shape of the proton-proton correlation function is affected by the nature of the final-state interactions.
The anti-correlation at low $q$ is a result of Coulomb repulsion.
More importantly, there is a characteristic final $^{1}$S$_{0}$ interaction peak at a relative momentum of 20~\mevc.  If there are two distinct emission timescales, a fast dynamical emission and a slow statistical emission, the height of this final interaction peak should primarily reflect the fraction of protons emitted during the fast pre-equilibrium stage of the reaction and the size of the emitting source. The width of the peak at $20$ ~\mevc~ is solely affected by the space-time of the fast pre-equilibrium source~\cite{verde02}. Therefore a detailed study of the overall shape of the correlation function allows one to extract the space-time extent of the source and constrain the relative contributions from fast and slow proton emitting components.

Experimentally the correlation function can be written as
\begin{equation}
  C(p_1,p_2) = \mathcal{N} \frac{A(p_1,p_2)}{B(p_1,p_2)}.
  \label{cor_dist}
\end{equation}
Here, the numerator from Eq.~\ref{cor_dist} is the distribution of two protons with momentum $p_1$ and $p_2$ detected in the same event. The denominator describes the uncorrelated background distribution and is constructed using so-called {\it event-mixing} method~\cite{kop74,lisa91} where each particle within a pair comes from a different event, taking into account the experimental two proton detection efficiency.   $\mathcal{N}$ is a normalization factor, which typically results in correlation functions that are close to unity at large relative momenta~\cite{Adams:2004yc,Aggarwal:2010bw}.

We used two different methods to extract the sizes of the sources presented in this paper.  In one method, we employed the imaging technique~\cite{Brown:1997ku,Brown:1997sn,Brown:2000aj} to extract both the size of the source and the source distribution profile $S(r)$ from the measured correlation functions.  In the other method, we  obtained source sizes by fitting experimental correlation functions with the Koonin-Pratt formula (Eq.~\ref{eq_kp}) assuming the Gaussian source distribution $S(r)$ given by%
\begin{equation}
  S(r) = \frac{\lambda_G}{(2\sqrt{\pi}R_G)^3} e^{-\frac{r^{2}}{4R_G^{2}}}
  \label{eq_gaus}
\end{equation}
For the Gaussian source, there are three free parameters: the normalization of the correlation function, $\mathcal{N}$, $\lambda_G$ and the source size $R_G$ parameters of Eq.~\ref{eq_gaus}.

According to Eq.~\ref{eq_source_integral}, $\lambda_G=1$
if the emission of all protons used to construct the correlation function is described by the source function.
While some protons are emitted over a short time scale after the collision and are strongly correlated, other protons can be emitted over very long timescales due to evaporation processes and secondary decays. Since the strength of the correlations reflects the spatial separation between the two protons at the time the second proton is emitted, early protons are not correlated with protons emitted at later times and late protons are only weakly correlated with each other. 

When both early and later emission occurs, the width of the peak in the \pp correlation function at 20~\mevc primarily reflects the early emitted particles (fast source with smaller source sizes).  Slowly emitted particles, coming from long-lived and more extended secondary decay sources primarily influence  the correlation function at low $q$-values~\cite{verde02}. If one is not primarily concerned with low $q$-values, these late emissions of protons largely reduce the magnitude of the correlation function while not usually strongly modifying its shape~\cite{verde02}.  In this case, Equation~\ref{eq_source_integral} has a more general form that reflects the fact that not all protons are correlated with each other, given to a good approximation by
\begin{equation}
\label{eq_source_integral_lambda}
4\pi \int S(r)r^2 dr = \lambda.
\end{equation}

The $\lambda$ parameter represents the fraction of pairs where both protons are emitted by the fast source represented by $S(r)$ over the range of $r$ represented in the integral in Eq.~\ref{eq_source_integral_lambda}. The remainder $1-\lambda$ contains the contributions from pairs at large separation $r$ outside of this range, where either one or both protons are emitted by the slow source at the late secondary decay stage of the reaction. The relevant proton pair fraction comes from the fast source; thus, $\lambda$, can be well approximated by~\cite{verde02}
\begin{equation}
\label{eq_fraction}
\lambda = f^2
\end{equation}
where $f$ and the remainder $1-f$ are the fractions of the total protons yields produced by the fast source and the slow source, respectively.

To minimize apriori assumptions about the source function we
follow the imaging techniques described in Ref.~\cite{Brown:1997ku,Brown:1997sn,Brown:2000aj,verde02,verde03}, and describe the source function $S(r)$ by an expression involving  three positive definite spline functions, which decreases monotonically with radius.  We take the half-width half-maximum of the extracted source profile, $r_{1/2}$, as a measure of the spatial extent of the source. This provides a simple size parameter that can be easily calculated even in the case of non-Gaussian source profiles where a $R_G$ parameter as in Eq.~\ref{eq_gaus} cannot be defined. In the specific case of a Gaussian source, the relation between $r_{1/2}$ and the size of the Gaussian source distribution (Eq.~\ref{eq_gaus}) is given by
\begin{equation}
  r_{1/2}=2\sqrt{ln2} R_{G}.
  \label{eq_r12}
\end{equation}
\section{Experimental Details}
We performed an experiment at the National Superconducting Cyclotron Laboratory (NSCL), where primary beams of \nuc{40}{Ca} (\nuc{48}{Ca}) with an E/A=80 MeV impinged on \nuc{40}{Ca} (\nuc{48}{Ca}) targets in a form of a thin mono-isotopic metallic foils of isotopic purity of about 97\% (92.4\%) by mass. We mounted the target near the center of the 4$\pi$ Array, which housed 215 fast/slow phoswiches covering 85\% of 4$\pi$ solid angle around the target in the laboratory reference frame. The 4$\pi$ Array, with an inside diameter of nearly 2 m, was instrumental in selecting central events by requiring a high transverse energy, $E_{t}=\sum_{i} E_{i}sin^{2}(\theta_{i})>150$ MeV~\cite{phair92}. Here, $\theta_{i}$ and $E_{i}$ correspond to the angles and energies of charged particles detected in the 4$\pi$ detector array. Assuming the transverse energy to monotonically decrease with impact parameter, this gate on $E_{t}$ corresponds approximately to an impact parameter range of $0<b~(fm)<4$.

In order to achieve both precise angular and energy measurements of the protons, required for correlation functions, we replaced one of the forward hexagonal modules of the 4$\pi$ Array with the High Resolution Array (HiRA)~\cite{hira07}. In our setup, HiRA consisted of 17 individual telescopes in a hexagonal configuration, each housing a 65$\mu m$ thin single-sided silicon strip detector followed by a 1.5 mm thick double-sided Si strip detector with each face having 32 strips with a pitch of 2 mm and an active area of 62.3 x 62.3 mm$^{2}$. The orthogonal orientation of the front and back strips of the thick Si detector, which was located ~63 cm from the target, allowed for angular resolution of $\delta\theta \approx 0.05^{\circ}$.  In order to allow the high-precision angular determination of the position, we measured the position of the target and silicon strips in HiRA with the Laser Based Alignment System (LBAS)~\cite{rog11}. Additionally, the Si detectors were backed by a cluster of four 39 mm long CsI(TI) crystals which served as the calorimeters. For this paper, we only analyzed protons which stopped in the CsI(TI) crystals.  This resulted in a proton momentum range of approximately 200-500~\mevc.
The angular coverage of HiRA with respect to the beam was $18< \theta_{Lab}~(deg)<58$ in the laboratory frame and $30<\theta_{CM}~(deg)<110$ in the center-of-mass frame.

%
\begin{figure}[ht!]
  \includegraphics[width=.5\textwidth]{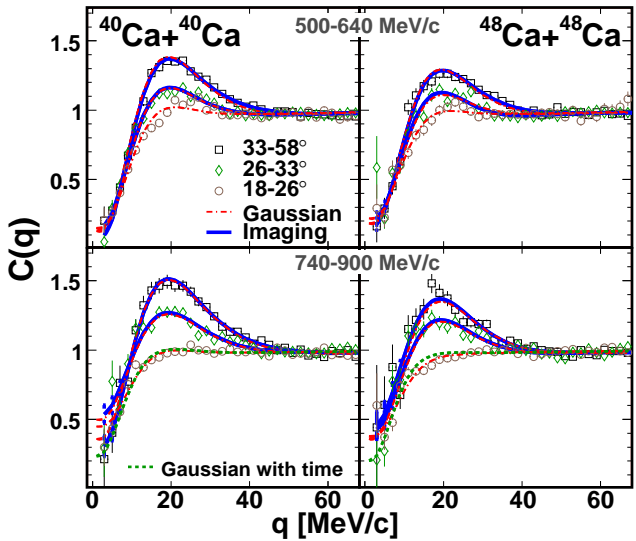}
  \caption{\label{fit_cor}
    Experimental correlation functions from \nuc{40}{Ca}+\nuc{40}{Ca} (left) and \nuc{48}{Ca}+\nuc{48}{Ca} (right). The upper panels include protons with low total momentum of the pair (500-640~\mevc) while the lower panels represent proton pairs with a high total momentum (740-900~\mevc).
    The dashed-dotted lines represent the results of the fit assuming the Gaussian source distribution.
    The solid lines are reconstructed correlation functions from imaging.
    The dashed lines represent the calculations assuming the Gaussian source distribution with non-zero lifetime (Eq.~\ref{eq_gaustime}); see Sec.~\ref{results} for more details.}
\end{figure}
\section{Experimental Results}
\label{results}

The correlation functions measured in our experiment are shown in Fig.~\ref{fit_cor}. The left panels present results from \CaFCaF\ and the right panels from \CaFECaFE\ collisions.  The upper and lower panels are  for protons with total momentum of the pair in the  laboratory frame of 500-640~\mevc\  and 740-900~\mevc, respectively.  The correlation functions at the most  backward angles ($33-58^{\circ}$) in the laboratory frame are represented by squares and at  forward angles ($18-26^{\circ}$) are shown as circles. The results at intermediate angles ($26-33^{\circ}$) are plotted as diamonds.

\begin{figure}[ht!]
  \includegraphics[width=.5\textwidth]{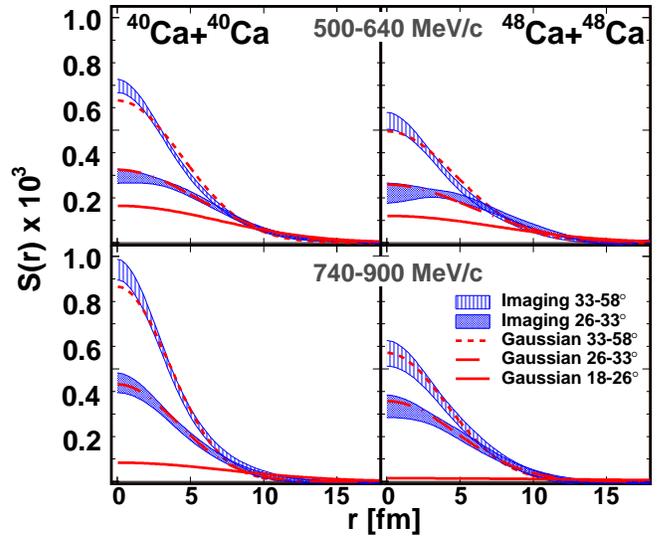}
  \caption{\label{source}Comparison of imaging technique to Gaussian fit of \pp correlation functions for \nuc{40}{Ca}+\nuc{40}{Ca} (left) and \nuc{48}{Ca}+\nuc{48}{Ca} (right). The upper panels include proton pairs with low total momentum (500-640~\mevc) while the lower panels represent proton pairs with a high total momentum (740-900~\mevc).}
\end{figure}

In order to get quantitative information about the proton emitting source we use the imaging technique to extract the imaged source function. The fits to the correlation function are shown as the solid lines in Fig.~\ref{fit_cor}. The corresponding extracted source functions are presented as the light cross-hatched and dark solid bands in Fig.~\ref{source} for $33-58^{\circ}$ and $26-33^{\circ}$, respectively.  In general, the correlation functions at backward angles have source functions that are larger and more localized around $r=0~fm$.

Imaging allowed us to reconstruct source distributions only at  backward and intermediate angles. The imaging technique fails at forward angles when the peak at $q=20$~\mevc\ is not well defined.
If the source were Gaussian, the peak would become negligible for large values of the $R_G$ parameter in Eq.~\ref{eq_gaus}, e.g. $R_G>5-6~fm$. Both the presence of sources with such large spatial extensions, and large statistical errors in the correlation function make convergence of the imaging method difficult to achieve at forward angles.

The fit quality and the normalization of the reconstructed source distribution, $\lambda_I$ from Eq.~\ref{eq_source_integral_lambda}, are given in Table~\ref{tab:results}.
If no constraints are placed on the shape of the source function, the imaging method can provide other solutions i.e. source functions $S(r)$, with comparably small values of $\chi^2/dof$, where $dof \approx 30$ is the number of data points minus the number of fit parameters. However, some of those solutions have unphysical properties, such as $S(r)<0$ so we exclude  them from the analysis and the error estimation of the $\lambda_I$ parameter.

We also performed fits to the experimental correlation functions using Eq.~\ref{eq_kp} and  assuming the Gaussian source function given by Eq.~\ref{eq_gaus}.
The corresponding fits to the correlation functions are denoted by the dashed-dotted curves in Fig.~\ref{fit_cor}. For the angular ranges of $\theta=26-33^{\circ}$ and $\theta=33-58^{\circ}$, these fits are nearly indistinguishable from the fits obtained via the imaging procedure, the latter shown as thick lines in Fig.~\ref{fit_cor}.
In these fits there are three fitting parameters: 1) the {\it size} of the source, $R_G$; 2) the $\lambda_G$ parameter (from Eq.~\ref{eq_gaus}); and 3) the normalization of the correlation function, $\mathcal{N}$ (from Eq.~\ref{cor_dist}). The best fit parameters are presented in Table~\ref{tab:results}.
The source distributions obtained from the Gaussian fit are plotted as the solid, dashed-dotted and dashed lines for $18-26^{\circ}$, $26-33^{\circ}$ and $33-58^{\circ}$ in Fig.~\ref{source}, respectively.
%
%
\begin{table*}
  \centering
  \medskip
  \begin{tabular}{| c | c | c | c | c | c | c | c | c | c | c | c | c | }
    \hline
    System & P & Angle & \multicolumn{5}{|c|}{Gaussian fit} & \multicolumn{4}{|c|}{ Imaging} & BUU \\
    \cline{4-13}
    &  [$\mevc$] &  [$^{\circ}$] & $R_{G}$ [fm] & $r_{1/2}$ [fm] & $\lambda_G$ & $f_G$ &  $\chi^{2}/{dof}$ & $r_{1/2}$ [fm] & $\lambda_I$ & $f_I$ &   $\chi^{2}/{dof}$  & $r_{1/2}$ [fm] \\
    \hline
    \hline
    \CaFCaF & 500-640 & 33-58 & $ 3.12^{+0.12}_{-0.06}$ & $ 5.20^{+0.21}_{-0.11} $ & $0.86^{+0.06}_{-0.04}$ & $ 0.93^{+0.04}_{-0.03}$ & 1.48 & $ 4.49^{+0.23}_{-0.51} $ & $ 0.93^{+0.13}_{-0.11}$ & $ 0.96^{+0.07}_{-0.06}$ & $ 1.48 $ & $5.29\pm0.10$  \\
            &  & 26-33 & $ 3.88^{+0.07}_{-0.08}$ & $ 6.46^{+0.12}_{-0.14} $ & $0.84^{+0.04}_{-0.04}$ & $ 0.92^{+0.03}_{-0.03}$ & 1.05 & $ 6.85^{+0.40}_{-0.47} $ & $ 0.85^{+0.14}_{-0.13}$ & $ 0.92^{+0.08}_{-0.07}$ & $ 1.18 $ & $6.24\pm0.09$  \\
            &  & 18-26 & $ 4.87^{+0.19}_{-0.15}$ & $ 8.11^{+0.31}_{-0.28} $ & $0.84^{+0.04}_{-0.04}$ & $ 0.92^{+0.03}_{-0.03}$ & 2.12 & $ -                    $ & $-$                       & $-$           & $-$        & $7.08\pm0.10$  \\
      \cline{2-13}
            & 740-900 & 33-58 & $ 2.52^{+0.17}_{-0.12}$ & $ 4.20^{+0.29}_{-0.21} $ & $0.61^{+0.11}_{-0.08}$ & $ 0.78^{+0.08}_{-0.06}$ & 0.88 & $ 4.06^{+0.23}_{-0.40} $ & $ 0.69^{+0.19}_{-0.12}$ & $ 0.83^{+0.11}_{-0.08}$ & $ 1.06 $ & $4.25\pm0.09$ \\
             &  & 26-33 & $ 2.91^{+0.22}_{-0.17}$ & $ 4.85^{+0.37}_{-0.29} $ & $0.48^{+0.12}_{-0.07}$ & $ 0.69^{+0.10}_{-0.06}$ & 1.28 & $ 4.71^{+0.40}_{-0.48} $ & $ 0.52^{+0.17}_{-0.10}$ & $ 0.72^{+0.12}_{-0.07}$ & $ 1.58 $ & $4.76\pm0.09$ \\
             &  & 18-26 & $ 5.40^{+0.41}_{-0.34}$ & $ 8.99^{+0.69}_{-0.57} $ & $0.54^{+0.05}_{-0.05}$ & $ 0.73^{+0.04}_{-0.04}$ & 1.52 & $ -                             $ & $-$                       & $-$   & $-$          & $5.33\pm0.10$ \\
      \hline
      \hline
     \CaFECaFE & 500-640 & 33-58 & $ 3.37^{+0.10}_{-0.09}$ & $ 5.62^{+0.17}_{-0.16} $  & $0.81^{+0.07}_{-0.06}$ & $ 0.90^{+0.04}_{-0.04}$ & 1.08 & $ 4.94^{+0.26}_{-0.54} $ & $ 0.84^{+0.17}_{-0.14}$ & $ 0.92^{+0.10}_{-0.08}$ & $ 1.41 $ & $5.69\pm0.11$ \\
               &  & 26-33 & $ 4.11^{+0.12}_{-0.16}$ & $ 6.85^{+0.21}_{-0.27} $  & $0.80^{+0.06}_{-0.06}$ & $ 0.89^{+0.04}_{-0.04}$ & 1.08 & $ 8.35^{+0.66}_{-0.73} $ & $ 0.81^{+0.16}_{-0.12}$ & $ 0.90^{+0.09}_{-0.07}$ & $ 1.17 $ & $6.81\pm0.09$ \\
               &  & 18-26 & $ 5.25^{+0.34}_{-0.36}$ & $ 8.74^{+0.52}_{-0.61} $  & $0.77^{+0.06}_{-0.06}$ & $ 0.88^{+0.04}_{-0.04}$ & 0.97 & $ - $ & $-$                       & $-$  & $-$       & $7.79\pm0.11$ \\
      \cline{2-13}
              & 740-900 & 33-58 & $ 2.85^{+0.18}_{-0.16}$ & $ 4.75^{+0.31}_{-0.28} $ & $0.59^{+0.12}_{-0.11}$ & $ 0.77^{+0.08}_{-0.08}$ & 1.27 & $ 4.69^{+0.52}_{-0.43} $ & $ 0.60^{+0.16}_{-0.11}$ & $ 0.77^{+0.10}_{-0.07}$ & $ 1.51 $ & $4.58\pm0.10$ \\
               &  & 26-33 & $ 3.34^{+0.15}_{-0.24}$ & $ 5.90^{+0.26}_{-0.41} $ & $0.59^{+0.08}_{-0.09}$ & $ 0.77^{+0.06}_{-0.07}$ & 1.47 & $ 5.85^{+0.70}_{-0.74} $ & $ 0.58^{+0.22}_{-0.13}$ & $ 0.76^{+0.14}_{-0.09}$ & $ 1.15 $ & $5.11\pm0.14$ \\
               &  & 18-26 & $ 9.83^{+5.21}_{-2.58}$ & $ 16.37^{+8.68}_{-4.30}  $  & $0.64^{+0.36}_{-0.09}$ & $ 0.80^{+0.23}_{-0.06}$ & 1.04 & $ - $  & $-$                       & $-$   & $-$        & $5.80\pm0.10$ \\
    \hline
  \end{tabular}
  \caption{Comparison of system size, angular and momentum dependence of  results obtained from reconstructed source distribution with imaging method, Gaussian fitting procedure and BUU transport model simulations.}
  \label{tab:results}
\end{table*}

The correlation functions reconstructed from imaging and obtained from the Gaussian fit are very similar and match the data well at most angles, as it is shown in Fig.~\ref{fit_cor}.  For the lowest momentum gate at $\theta=33-58^{\circ}$, the peaks in correlation functions for the Gaussian sources are narrower and their tails lie consistently below the data and the imaging results for $q\approx 40$~\mevc.  This gives rise to the slightly wider widths of the corresponding Gaussian sources shown in Fig.~\ref{source} for these data. For the other gates, the results for Gaussian and imaging analyses are very similar;
in some cases, the source functions provided by imaging method  are slightly more localized at $r=0~fm$ than the corresponding fits with the Gaussian source functions.
At the most forward angles where the size of the source is large and the correlation effect is not as strong as in the experimental data collected at backward angles, it was not possible to constrain the source function adequately via the imaging technique. There we used the more constrained Gaussian source function in order to extract information about the space-time extent of the source. Fortunately, the similarity between Gaussian and imaging analyzes at the other angles provides support for us to use the Gaussian approach and lends confidence to the information it provides.

To provide the simplified measure of the source, we characterize the extracted sources using $r_{1/2}$ (also used in e.g.~\cite{verde02,verde03,Chung:2002vk}) for each set of data and method used to extract the source distribution or its size.  Results are presented in Table~\ref{tab:results} for both reaction systems, both pair momentum ranges in the laboratory frame, and all three angular selections. With the exception of the lowest momentum gate at $\theta=33-58^{\circ}$, the values for  $r_{1/2}$ are  consistent between imaging and a Gaussian fit.
\begin{figure}
  \includegraphics[width=.5\textwidth]{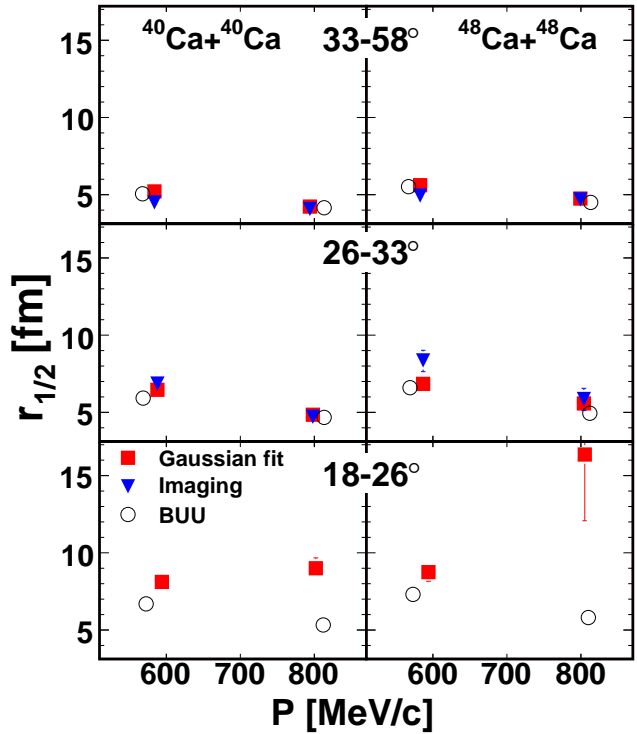}
  \caption{\label{srrmax} $r_{1/2}$\ as a function of total momentum for \CaFCaF\ and \CaFECaFE\ collisions  and all three angular ranges.  The sizes of source from data using the imaging technique are given by red closed circles while those from the Gaussian technique are shown as blue closed triangles.  Source sizes from BUU are shown as black open circles.}
\end{figure}

The sources from the collisions with larger initial geometry, \nuc{48}{Ca}+\nuc{48}{Ca} (N/Z=1.4), are systematically larger than those from \nuc{40}{Ca}+\nuc{40}{Ca} collisions (N/Z=1).  The average increase in source size with A somewhat exceeds $A^{1/3}$, which suggests that the average freezeout density is somewhat lower for the \nuc{48}{Ca} than for \nuc{40}{Ca}. Due to the large value of the neutron-proton cross section which significantly exceeds the \pp cross section, the relevant density for proton freezeout may be the neutron density rather than the total nuclear density.   In this case, the additional neutrons in the \nuc{48}{Ca}+\nuc{48}{Ca} may shift the freezeout to lower overall density. Such a shift reflects detailed differences in the transport of neutrons and protons that could be used to extract information about the relevant
in-medium cross section. Calculations indicate, however, that such effects are subtle and dwarfed by the qualitative difference between the sizes at forward and backward angles and we, therefore,  defer such detailed model investigations to a latter publication.

Clearly, the strong angular and momentum dependence of the extracted source size is a much more dramatic trend. The observed large increase in the source size occurs at forward angles at velocities comparable to that of the beam. (The total momentum of two beam velocity protons is approximately 800 MeV/c.) Correlation functions of similar magnitudes have been reported for protons evaporated from heavy residues produced in \nuc{129}{Xe}+\nuc{27}{Al} reactions~\cite{Lisa:1994zz}, and for protons emitted at energies comparable to the Coulomb barrier in \nuc{40}{Ar}+\nuc{197}{Au}~\cite{Pochodzalla:1987zz} and  \nuc{129}{Xe}+\nuc{197}{Au} reactions~\cite{Verde:2007ma}. In the latter case, however, a relatively small fraction of fast protons (f=0.30) was reported.
From the $\lambda_G$ and $\lambda_I$ parameters obtained from the Gaussian fit and imaging method 
we calculated the fraction $f$ of short time scale emitted protons~\cite{verde02},  according to Eq.~\ref{eq_fraction}.
The results are summarized in Table~\ref{tab:results}.
~We show that the values of this $f$ parameter are consistent between both methods.
All proton fractions exceed 0.5 and there is very little momentum or angular dependence of the $f$ parameter. It is interesting that the large sources at $\theta=18-26^{\circ}$ and total momenta of $740-900$~\mevc~ observed for both reactions also have relatively large fast fractions eg. $f > 0.7$. This implies that more than 70\% of the two-proton emission occurs at relative separations of $r<15$ fm. This appears to exclude significant contributions ($>$30\%) from evaporative emission at emission time delays much greater than about 150 fm/c. 

On shorter timescales, the relative importance of a spatial expansion of the projectile remnant versus the effects of an extended lifetime cannot be distinguished with angular averaged correlation functions such as those presented here. However, some information can be gleaned by considering the limits of an expansion followed by an instantaneous emission versus an emission that extends over timescale of the order 10's of fm/c. For the limit of instantaneous emission, we approximate the corresponding density by assuming that it is uniform with same RMS radius as the Gaussian distribution given by the best fit. In this approximation, the freezeout density would be approximately
$\varrho_{fre} \approx A_{spec}/\{\frac{4\pi}{3} (\sqrt{5}R_{G})^{3}\}$.
Assuming, the projectile contains $A_{spec}=20$ nucleons prior to fragmentation and $R_{G}=3-9$ fm, one obtains estimates for $\varrho_{fre}$ of $\varrho_{fre}=0.004-0.1 \rho_{0}$. This is somewhat below the density range assumed by statistical simultaneous multi-fragmentation models~\cite{Bondorf:1995ua,Gross1997119}.
It is also below the density range, $\rho = 0.2 - 0.4 \rho_{0}$, extracted from $d-\alpha$ correlations for the participant source in \nuc{129}{Xe}+\nuc{197}{Au} collisions~\cite{Verde:2007ma}. Both comparisons suggest that the source fragments over a non-zero timescale.

Alternatively, we assume that decay occurs from a spherical source with $R_{G}$ and vary the timescale of the decay. Following Koonin~\cite{koon77} we assume a Gaussian emission time distribution: i.e.  emission rate $\propto exp(-t^{2}/\tau^{2})$. This leads to a source function of the form:
\begin{equation}
  S(r) = \frac{\lambda_G}{(4\pi)^{3/2}R_G^2 \sqrt{R_G^2 + 0.5(v\tau)^2}} e^{- \frac{r_{\perp}^{2}}{4R_G^{2}} -  \frac{r_{||}^{2}}{4(R_G^{2}+ 0.5(v\tau)^2)} }.
  \label{eq_gaustime}
\end{equation}
Here, $v=|\overrightarrow{v}|$, where $\overrightarrow{v}=\overrightarrow{V}-\overrightarrow{V_{0}}$ is the magnitude of the velocity $\overrightarrow{V}$ of the center of mass of the two protons relative to the velocity $\overrightarrow{V_{0}}$ of the source, $r_{\perp}$ ($r_{||}$) is the component of $\vec{r}$ perpendicular  (parallel) to $\vec{v}$.

The beam momentum per nucleon is roughly equal to the average proton momentum
for the data with $18^{\circ}<\theta<26^{\circ}$ and $740 \leq P~(MeV/c)<900$.
Thus, most of the protons within this gate must be preferentially emitted perpendicular to the beam leading to an estimated velocity of  $v \approx 0.16 c$.
In this scenario, the space-time extent measured for those particles  is a combination of the spatial dimension ($R_G$) and the lifetime of the source ($\tau$).
The dashed lines in the lower panels of Fig.~\ref{fit_cor} correspond to the correlation functions obtained with the source distribution from Eq.~\ref{eq_gaustime}, where $R_{G}=3$ fm  and $\tau=100$~fm/c ($\approx 3.3 \times 10^{-22}$s)  for \nuc{40}{Ca}+\nuc{40}{Ca}  and $R_{G}=3.5$ fm and $\tau=135$~fm/c ($\approx 4.5 \times 10^{-22}$s)
\nuc{48}{Ca}+\nuc{48}{Ca} reactions.

Those calculations show reasonable agreement with the experimental correlation functions.  This "lifetime" is relatively short for a statistical evaporation process, but comparable to the times for expansion and disassembly during a multifragmentation process~\cite{Kim:1991zz,Bowman:1993zz,Kampfer:1993zz,Fox:1993zz,Bauge:1993zz,Cornell:1995zz}. 

To illustrate the inconsistency of the large source sizes at forward angles with a straightforward dynamical origin, we simulated \CaFCaF\ and \CaFECaFE\ collisions at E/A=80 MeV.  We chose a parametrization of BUU such that an energy dependent in-medium nucleon-nucleon cross section reduction was employed~\cite{dan02}.  We also included momentum dependence in the mean field with a soft equation of state~\cite{dan00}.  We chose the density dependence of the symmetry energy to be $\gamma=0.7$, which is in agreement with Ref.~\cite{tsang09}.  We also included the production of $A\leq$3 clusters~\cite{dan91,dan92}; this tends to increase $r_{1/2}$ by approximately 1 fm. From the information provided by the transport model we constructed the source functions for the same momentum and particle emission angle in the laboratory as used in the experimental analysis.  We included only the protons emitted at energies and angles that could have been detected in the experiment. We calculated the quantity $r_{1/2}$ from the source distribution. We show the comparison between theoretical and experimental values for $r_{1/2}$ in Fig.~\ref{srrmax}.

In Fig.~\ref{srrmax}, we can see that BUU can reproduce the experimental data well at backward and intermediate angles for both pair momentum ranges measured in the laboratory frame,  but underpredicts the sizes  at forward angles, especially for protons in the high momentum gate. These high momentum particles move at close to the beam velocity. We have calculated source radii for a wide variety of different mean fields and nucleon-nucleon cross sections, but have not been able to find a choice of transport parameters that result in significantly larger source radii at forward angles and beam velocity. Such large radii indicate emission from a source that is much larger or longer-lived or both compared to the source that can be predicted by a dynamical model such as the BUU approach. A long lived source could explain the discrepancy with the BUU calculations, however a very long-lived source is inconsistent with the large fast fractions $f>0.7$ deduced from our measurements. However, BUU suppresses many fluctuations that lead to rapid multifragment disassembles. The failure of the BUU to describe the p-p correlations at forward angles and beam velocities provides a clear demonstration of the importance of such processes in this kinematic domain.

\section{Conclusions}
We studied the angular and momentum dependence of \pp correlations for central \nuc{40}{Ca}+\nuc{40}{Ca} and \nuc{48}{Ca}+\nuc{48}{Ca} nuclear reactions at E/A=80 MeV. We found a strong angular dependence within \pp correlation functions reflecting the different space-time extent of the source selected.  Sources observed at backward angles, in the laboratory frame, reflect the participant zone of the reaction, while much larger sources are seen at forward angles  are dominated by expanding, fragmenting and evaporating projectile-like residues. The obtained results show a decreasing source size with increasing momentum of the proton pair emitted at backward and intermediate angles. In contrast we observe a weak trend in the opposite direction at forward angles.
At some level, these trends are consistent. In the rest frames of the respective sources, higher velocity protons are more strongly correlated than their lower velocity counterparts, consistent with emission from expanding and cooling sources. The protons with small laboratory momenta at backward angles move slowly in the rest frame of the participant source. Protons with large laboratory momenta at backward angles move rapidly  in the rest frame of the participant source. In contrast, the highest momentum protons at forward laboratory angles are nearly at rest relative to the fragmenting projectile remnants, and the lower momentum protons at forward angle are actually moving at a higher relative velocity to the fragmenting projectile remnants. 
In both angular domains, we therefore observe smaller sources for protons moving at higher velocities in the frame of the source.

Long evaporation times are not consistent with the fast fractions extracted from the correlation functions forward angles. The time scales estimated from our correlation functions are consistent with bulk multi-fragmentation time scales that have been extracted by fragment-fragment correlation functions.

We find that BUU transport calculations reproduce the data well at backward and intermediate angles, but underpredict to reproduce the source sizes at forward angles at high momentum. There the data are consistent with expansion, multi-fragmentation and subsequent evaporation. The failure of the BUU to reproduce the source functions for this case can be attributed to the suppression of the fluctuations leading to multi-fragmentation in this approach.

In all cases, the \nuc{48}{Ca}+\nuc{48}{Ca} reaction system results in larger sources than the \nuc{40}{Ca}+\nuc{40}{Ca} reaction system, which can be partly attributed to a sensitivity of the source distribution to the initial size of the projectile and target nuclei. 
However, the effect appears to be somewhat larger than the $A^{1/3}$ scaling expected from such geometrical arguments.


\section{Acknowledgments}

We would like to thank S. Pratt and D. A. Brown for their help with the imaging process.  
We wish to acknowledge the support of Michigan State University, the Join Institute for Nuclear Astrophysics, the National Science Foundation Grants No. PHY-0216783, No. PHY-0606007, No. PHY-0822648, and No. PHY-0855013, and the U.S. Department of Energy, Division of Nuclear Physics Grant No. DE-FG02-87ER-40316 and Contact No. DE-AC02-06CH11357.

\end{document}